# MODEL DRIVEN WEB APPLICATION DEVELOPMENT WITH AGILE PRACTICES

Gürkan Alpaslan [1] and Oya Kalıpsız [2]

[1,2] Department of Computer Engineering, Yıldız Technical University, Istanbul, Turkey

## ABSTRACT

*Model driven development is an effective method due to its benefits such as code transformation, increasing productivity and reducing human based error possibilities. Meanwhile, agile software development increases the software flexibility and customer satisfaction by using iterative method. Can these two development approaches be combined to develop web applications efficiently? What are the challenges and what are the benefits of this approach? In this paper, we answer these two crucial problems; combining model driven development and agile software development results in not only fast development and easiness of the user interface design but also efficient job tracking. We also defined an agile model based approach for web applications whose implementation study has been carried out to support the answers we gave these two crucial problems.*

## 1. INTRODUCTION

Model driven development or MDD is a method; proposes to produce the source codes via models [1], [2]. Models are the abstracted representation of the system elements [3]. Created models are transformed to source codes by MDD tools with automated code generation property [4], [5]. This is very beneficial attribute, since it reduces the human factor on software coding. In other words, it leaves the coding part only to computers. Thus, developers only focus on creating the system models properly [6].

Models are utilized for web applications on different methodologies like WebML [7], [8], UWE [9], [10] and OOHDM [11]. These methodologies are mostly based on Unified modeling Language. Another method is Mockup driven development which is based on prototyping the web applications [12]. Mockups are the dynamic user interface prototypes created by mockup development tools [13], [14]. These tools provide to transform the mockups to executable web pages created by Hyper Text Markup Language (HTML), Cascade Style Sheet (CSS), JavaScript codes and other web development technologies [15]. The critical advantage of using tools than hand-coding is to utilize the last technologies, low error rate and pace.

On the other hand, agile practices [16], [17] aim to deliver executable software quickly. Agile based development methods do not consider the documentation and the structure of them are iterative; software is developed in pieces. This structure provides more flexible skeleton and responses the feedbacks better during the life cycle [18].

In literature, there are some life cycle diagrams combining the agile practices on model driven development [19]. One of the prior studies in literature is the agile model driven development (AMDD) high level life cycle [20], [21]. It basically proposes a life cycle consists two main phases: inception phase and development phase. The inception phase is the general modelling part of the whole system. Iterations are implemented in the development phase.

Another life cycle is the Hybrid MDD development method [22]. It is a developed method of the AMDD high level life cycle. It consists two main phases like high level life cycle, but it defines the three development teams working parallel. These teams are called as model development team, agile development team and business analyst team, which are defined in Section 3.

There are also MDD SLAP method [23], which is developed by Motorola Company in order to work their own company agile projects, and Sage MDD [24] method, which is developed in order to use on developing multi agent systems. The methods are evaluated with the criteria of contribution and target platform they are developed for (Table 1).

Table 1. Main AMDD life cycles in literature

|  | **Main Contribution** | **Target Platform** |
| --- | --- | --- |
| AMDD High Level Life Cycle | First life cycle combining agile practices with model driven engineering | General projects |
| Sage MDD | Based on the integration of the different models incrementally and iteratively | Multi-agent systems |
| Hybrid MDD | Described the parallel working teams on AMDD High Level Life Cycle | Small or medium size general projects |
| MDD-SLAP | Identify the relation between agile principles and model-driven practices and implemented on Scrum method. | Telecommunication systems |

To evaluate the cost estimation of web projects have different properties than others [29]. Generally, web projects have small team groups and not trustable to evaluate the project with code line numbers. Instead of that, the main factors which are personnel, product, platform and project factors have been utilized for better estimation. As a result of estimation, the result give the cost tendency of method (Section 4.2).

The main contributions of this work are: (1) to provide an agile supporting model-driven approach customized for web applications, (2) to propose a life cycle, implemented on real projects for developers; (3) and to provide the cost evaluation of the approach.

The paper is structured as follows: in Section 2, we describe our proposed approach in detail and then in Section 3, we detail the implementation work of approach. Section 4 describes the discussions and results of the work and finally, in Section 5 we draw some conclusions and present our future work on this field.

## 2. PROPOSED APPROACH

The approach is based on the Hybrid MDD [22] method. The Hybrid MDD method is customized for web applications. The dynamic prototyping method for web applications called mockup driven development [12] is utilized on the life cycle. We defined which parts are created by using

International Journal of Software Engineering & Applications (IJSEA), Vol.7, No.5, September 2016

mockups automatically and which parts are created by handcrafted coding. After all, these parts are integrated properly.

For web applications, system architecture can define as two programming parts called client-side and server-side [25]. Client-side codes are not generated on servers. Server-side codes are generated on server and send to clients. In our approach, we aim to coding all client-side parts via model transformation. Client-side part is developed from models with automated code generation and server-side part is developed by handcrafted codes. Finally, these parts are integrated and final software is emerged.

## 2.1. The Steps of the Approach

The approach is progressed iteratively and incrementally. Three teams are worked parallel throughout the life cycle. These teams are the model driven development team, agile development team and business analyst team. Model development team is responsible of the model infrastructure construction, creation of web models with their attributes and functions and the automated code generation parts. Agile development team is responsible of the test environment creation and handcrafted codes. Business analyst team is responsible of the interaction with customer, creation of the requirement of the system. These teams work with high cooperation as the result of agile principle. The steps of the method are basically illustrated (Fig. 1).

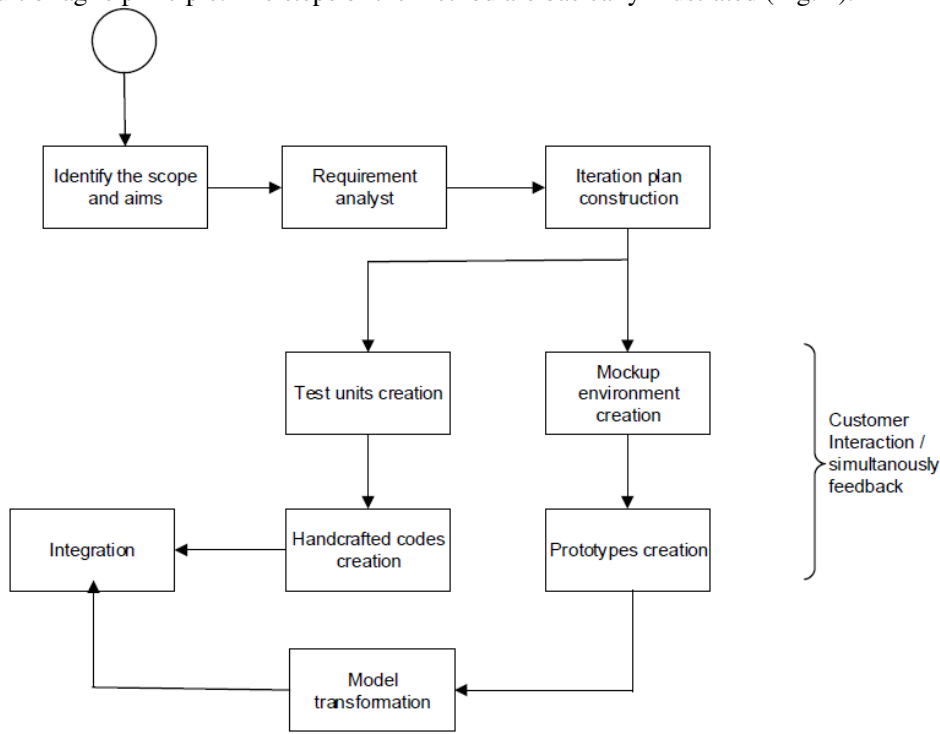

Figure 1. The main steps of the approach

The approach starts with identifying the system scope by business analyst team. The business analyst team describes the requirements by cooperation with customer. After the requirement analyst, all system is divided the iterations by the order of their priority. The development phase starts with most important priority iteration. After this point, all three teams work simultaneously. Agile development team creates the test units; while MDD team decides the proper model development tool and sets up; and business analyst team interacts with customer and coordinates all team members. After the iteration skeleton is completed, agile team develops handcrafted codes like database creation, database connection classes; model-driven development team creates the model with assigning their attributes and functions. In this process, the aim should be creation of all client-side codes by only model generation. Only server-side codes will be coded by handcrafted coding. After all, handcrafted coding parts and codes generated from model are integrated. These process repeats until the all iteration is completed.

### 2.2. Implementation of the Testing Phase

In the approach, test-driven development (TDD) [26] method is utilized. TDD proposed to produce the test units before the source code is created. In our approach, all of iterations begin with the creating the test units. After the integration part for all iterations, iteration test is implemented. When all the iterations are completed, integrated system is tested for compatibility and integrity.

### 2.3. Customer Role on the Approach

Customer interaction with the developers is a significant part of the approach. All the parts of the process, the iteration artifacts have to be presented to customer and feedbacks have to be evaluated immediately. Most of the parts of the process, pair development is proposed. For instance, model design is proposed to implement with a customer representation.

### 2.4. Agile Practices Utilization

Agile practices are applied to the structure of the approach. Agile modelling [27] is generally utilized for this purpose. Agile modelling is a method for modelling properly to the agile principles. The general utilization of the agile practices in our approach is clarified (Table 2).

Table 2. The utilization of the agile practices in the approach

| Agile Practices | Utilization in the Approach |
|---|---|
| Iterative and incremental development | The approach progresses iteratively and incrementally |
| Working software over comprehensive documentation | The models are used for documentation in the approach, instead of creating external documentation |
| Rapid feedback | All stakeholders work together with high interaction and responds to the feedbacks immediately |
| Continuous integration | Models are created part by part and regularly integrates |
| Test-driven development | All test units are developed before creating the source codes throughout the process |
| Pair programming | Mockups and requirements are created with customer representation |

## 3. CASE STUDY

### 3.1. Overview

The approach is implemented in Software Quality Research Lab in our university. For this purpose, two different teams have been constructed with different project subject.
Team 1 has developed the cinema ticket system; while team 2 is being developed the library registration system. For model development tool, Axure [28] tool has been utilized. Team members have the enough experience about both programming languages and basis software engineering methods.

### 3.2. The Progress of the Implementation

Firstly, our approach has introduced to the teams. Every week, with the meetings, the progress and coherency to the approach are checked. They have completed their projects in 18 weeks. The progress of the work is illustrated (Fig. 2).

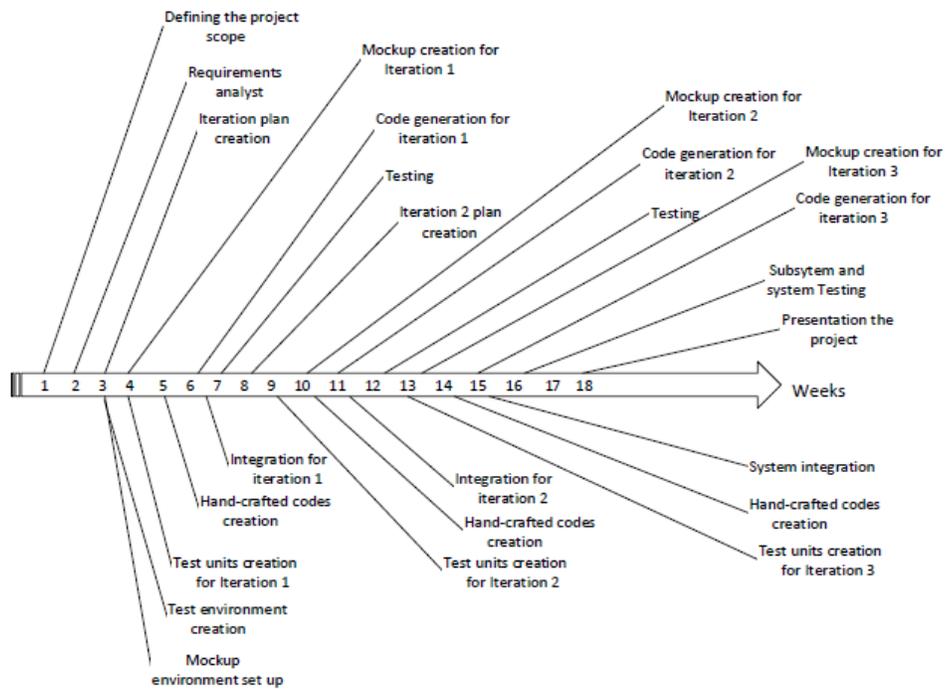

Figure 2. The progress of the case study projects

The projects are completed in three iterations. Generally, project progress can divide into two parts. First part is the planning part and second part is iteration parts which are repeated the same process in every iteration.

### 3.3. Testing Process of the Implementation Study

For the testing process, black box test technic is utilized. Black box testing is a method, proposed to test the working system structure rather that the code structures. The reason of the preferred this method is due to the biggest part of the codes has been created by automated code generation.

Generated codes have big complexity for normalizing and analyzing to test their structures. Besides, generated codes have fewer tendencies to error possibilities.

For both team projects, testing process is implemented in pieces. Firstly, general aspects are constructed for whole project basically for mapping the process. After that, testing units are created for every iteration as every iteration is being started. As the iteration artifact is constructed, iteration test is implemented. After three iterations, whole system is tested.

### 3.4. Challenges during the Implementation Study

One of the challenges teams must solve is the integration the handcrafted codes with generated codes created by automatically transformed from models. Teams have utilized two methods for solving this problem: (1) adding the handcrafted codes inside the generated codes with special tags, (2) creating the particular class for handcrafted codes and associating it with generated codes.

Another challenge is the complexity on the generated codes. Transformation tools create lots of line codes and it must be analyzed by developers for making changes. To decrease the complexity, tags are added on important model elements, hence, changings are followed more easily.

## 4. RESULTS AND DISCUSSION

### 4.1. Evaluation in the terms of Software Architecture

The proposed approach begins with the inception part where system is analyzed and the architecture is implemented. In the development phase, for all iterations, the analyst, design, coding and testing are implemented specially for that iteration. This process repeats as the number of the iteration. When all iterations completed, whole system is tested and all the iteration artifacts are integrated. The software development main processes which are analyst, design, coding and testing, are implemented with different orders throughout the approach. The total time efforts for those are approximately estimated. The proportions of the processes are illustrated (Fig. 3).

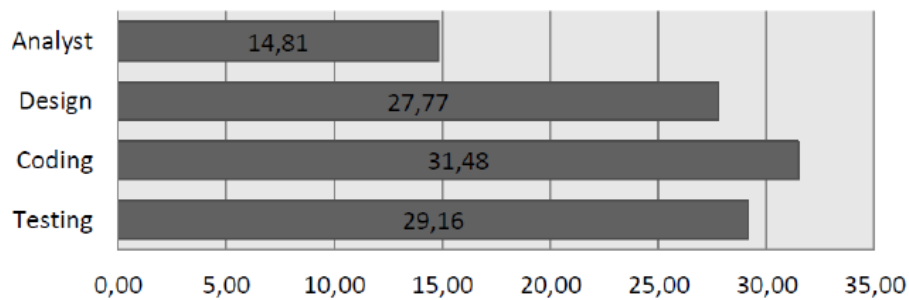

Figure 3. The proportion of the main software development processes on approach

Proposed approach is based on agile architecture skeleton, thus, analyst part is proportionally less than others. Coding part is included the handcrafted coding and coding from models. These proportions are given for three iteration included implementation study.

In addition, thanks to the three teams, which are business analyst, agile development and model-driven development teams, work simultaneously; so, production time is reduced. Approximately estimated 40% work load is implemented simultaneously in the implementation study.

## 4.2. Cost Estimation of the Approach

Cocomo II model [30] is utilized for cost estimation of the proposed approach. Cocomo II model includes mainly four categories which are personnel factors, product factors, platform factors and project factors. These factors contain totally 17 cost drivers whose has own scale ranges. These ranges are constraint values as described in Table 3. Nominal values are defined as 1.00 rate and the others are defined in regards to proportional of nominal. If the value is greater than 1.00, it means that state increases the cost effort. If the value is less than 1.00, the situation is effects positively for cost load.

Every cost driver in Cocomo II is evaluated for our approach. (Table 3) Effort adjustment factor is the value for evaluating the cost. Effort adjustment factor (EAF) is calculated by multiplied all rated values. (Eq. 1)

$$EAF = \prod_{i=1}^{n} X_i \quad (1)$$

In regarding of the ratings on Table 3, the cost estimation value is determined by multiples of all rates. So,

$$EAF = 1.11561 \quad (2)$$

The value is on the higher side of the nominal point, but the deviation score is not much. It is proved that approach is tendency to nominal line and has acceptable cost effort. For this approach, other criteria in Cocomo II like code lines are ignored, since the structure of the approach.

Table 3. Cost Effort Estimation in regarding to COCOMO II model

|  | Very Low | Low | Nominal | High | Very High | Our Rates |
|---|---|---|---|---|---|---|
| Required software reliability | .82 | .92 | 1.00 | 1.10 | 1.26 | 1.10 |
| Database size |  | .90 | 1.00 | 1.14 | 1.28 | .90 |
| Product complexity | .73 | .87 | 1.00 | 1.17 | 1.34 | 1.00 |
| Required Reusability |  | .95 | 1.00 | 1.07 | 1.15 | 1.07 |
| Documentation match to life cycle needs | .81 | .91 | 1.00 | 1.11 | 1.23 | 1.11 |
| Execution Time Constraint |  |  | 1.00 | 1.11 | 1.29 | 1.00 |
| Main Storage Constraint |  |  | 1.00 | 1.05 | 1.17 | 1.00 |
| Platform volatility |  | .89 | 1.00 | 1.15 | 1.30 | .87 |
| Analyst capability | 1.42 | 1.19 | 1.00 | .85 | .71 | 1.00 |
| Programmer Capability | 1.34 | 1.15 | 1.00 | .88 | .76 | 1.15 |
| Applications experience | 1.22 | 1.10 | 1.00 | .88 | .81 | 1.10 |
| Platform experience | 1.19 | 1.09 | 1.00 | .91 | .85 | 1.19 |
| Language and tool experience | 1.20 | 1.09 | 1.00 | .91 | .84 | 1.2 |
| Personnel continuity | 1.29 | 1.12 | 1.00 | .90 | .81 | .90 |
| Use of software tools | 1.17 | 1.09 | 1.00 | .90 | .78 | .78 |
| Multisite development | 1.22 | 1.09 | 1.00 | .93 | .86 | .86 |
| Required development schedule | 1.43 | 1.14 | 1.00 | 1.00 | 1.00 | 1.00 |

## 4.3. Evaluation in the terms of Security and Maintenance

Another best parts of MDD are security and maintenance benefits. MDD helps to reduce the human factor, meanwhile it is low error prone. This situation reduces the human based errors and makes the system development in an way that development tools control the process. In addition, with MDD, it is more flexible for change, so that, maintenance is easy as well. Changings are implemented on models and MDD reflects the model to real system.

## 4.4. The Strong Aspects and Weak Aspects of the Approach

We requested teams evaluating the strong aspects and weak aspects of the approach. Team 1 has defined the fast development and easiness of the user interface design as strong aspects of the approach. It has also defined the difficulty of the generated code analyst as weak aspect of the approach.

Team 2 has defined the regular job tracking as strong aspect of it. It has defined the complexity after the model transformation as weak aspect of it.

## 4.5. Research Questions

RQ1: What are the methods to integrate the model transformed codes with handcrafted codes?
After the mockups are transformed and agile team completes handcrafted codes, two methods are used for integration by teams. One of them is to add codes into the generated codes with using special tags. Other method is to create the handcrafted codes with using object classes and associates them on the common structure.

RQ2: What sizes of web projects are suitable for the approach? Is there any constraint?
The approach has developed on Hybrid MDD structure. Hybrid MDD is suggested for small or medium size projects. Customized version of that for web application, it does not require a constraint thanks to mockups eligible structure. However, big data causes more complexity for analyzing.

RQ3: What would be the main differences for the projects implemented in the industry ?
The main difference is the structure of the development team. In industry, team would be more experienced and organized on their areas. This is a factor that decreases the cost estimation. Another difference is the product size. Industry projects has more complexity as well as should be more flexible owing to customer factor. Customer's requests have more tendency to alteration in industry projects rather than university research projects.

RQ4: Is there any challenge about updating the previous created mockups?
Mockups are the visual and eligible structures, and currently mockup development tools are providing adequate features about it. In addition, model tagging [10] is proposed for decreasing the complexity and increase to intelligibility. Model tagging is a method to decrease the complexity of the model transformation. It proposes putting tags to every element in the model, so after the transformation it eases to track the codes.

RQ5: What are the main contributions of the approach ?
In this research, a life cycle, mainly based on the integration of the separately produced parts and prototyping method with agile support, is defined specially for web applications. The research provides to web developers an approach and document that can be utilized from the starting point of the project to software release.

## 5. CONCLUSIONS

In the paper, it is aimed to provide a life cycle for web application development. The proposed approach is based on the integration the client-size codes and server-size codes. To accelerate the process, parallel working teams are proposed. For this purpose, the life cycle of the Hybrid MDD method is utilized as skeleton. Model web prototyping method called Mockup has been implemented on the life cycle. As well as, agile practices are utilized for faster and flexible developing, better analyst process, rapid feedback and more. In addition, case study has been carried out by two different teams and obtained the feedbacks by participants.

Throughout the implementation study, the most significant challenge of the participants is the integration the client-side codes with server-side codes. Mockup driven transform tools have created the lots of code units than they expected. For basic web page, it has produced hundreds line HTML, CSS and JavaScript codes. The reason of that is the transform logic of the tools. They described every minimal element in the separated tags. In fact, the basic HTML, CSS script lines has caused not to analyze the meaning of the code. About this situation, our suggestion is to focus on the models instead of the codes. It has been a problem developing by models for participants who accustomed the developing by codes. Hence, this approach requires tracking regularly in order to implement properly.

Furthermore, the approach is evaluated positive by participants about the subject of visual design simplicity. User interface design has been created significantly fast by exclusive Mockup tools.
In the terms of cost factors is concluded as nearly nominal range. Effort adjustment factor is calculated as 1.11561 for case studies and this value is defined that the approach is in the acceptable range. This study has been implemented by intermediated level development teams, so that we can conclude that, by developing just the personnel factors, the effort can reach the nominal value.

For future work, it is planning to implement the approach on a big size projects. In the study, general perspective of the approach is illustrated and the applicability of that is proved for little or medium size projects. Moreover, reverse engineering for web application models is the interesting area for research.

## ACKNOWLEDGEMENTS

The authors would like to thank Software Quality Research Lab members for their support of this research. We are also grateful to all the practitioners for their participation and feedback.

**Authors**

**Gürkan Alpaslan** received the B.S. degree in computer engineering from Istanbul University in 2012 and M.S. degree in computer engineering from Yıldız Technical University in 2015. He is currently Ph.D. student and working as a research assistant at the Computer Engineering Department of Yıldız Technical University. His main research interests are software engineering and database systems.

**Oya Kalıpsız** received her M.S. degree in system analysis from Istanbul Technical University in 1984. She received her Ph.D. degree from Istanbul University in 1989 with a study on Hospital Information Systems. She is currently working as the member of Software Quality Research Lab in Computer Engineering Department of Yildiz Technical University. Her main research interests are software engineering, database systems, data mining, system analysis, and management information systems.